\documentclass[twocolumn,prl,showpacs]{revtex4}
\usepackage{graphicx}
\usepackage{dcolumn}
\usepackage{bm}
\usepackage{amsmath}

\begin{document}

\preprint{}
\title{Anomalous magnetoresistance on the topological surface}
\author{Takehito Yokoyama$^{1}$, Yukio Tanaka$^{2}$, and Naoto Nagaosa$^{1,3}$}
\affiliation{
$^1$ Department of Applied Physics, University of Tokyo, Tokyo 113-8656, Japan \\
$^2$Department of Applied Physics, Nagoya University, Nagoya, 464-8603, Japan \\ 
$^3$ Cross Correlated Materials Research Group (CMRG), ASI, RIKEN, WAKO 351-0198, Japan 
}
\date{\today}

\begin{abstract}
We investigate charge transport in two-dimensional ferromagnet/feromagnet junction on a topological insulator. 
The conductance across the interface depends sensitively on the directions of the magnetizations of the two ferromagnets, showing anomalous behaviors compared with the conventional spin-valve. 
This stems from the way how the wavefunctions connect between both sides.
It is found that the conductance depends strongly on the in-plane direction of the magnetization. Moreover, in sharp contrast to the conventional magnetoresistance effect, the conductance at the parallel configuration can be much smaller than that at the antiparallel configuration. 
\end{abstract}

\pacs{73.43.Nq, 72.25.Dc, 85.75.-d}
\maketitle



%

%



Spintronics aims to manipulate and/or use the spin degrees of freedom in device functions.
There are two mainstreams in spintronics: the control of charge transport by 
spins~\cite{Zutic,Baibich,Julliere,Maekawa,Moodera,Wolf}, and the control of spins  by the electric field~\cite{Datta,Nitta,DP,Murakami,Kato,Wunderlich}. 
In the former, giant magnetoresistance~\cite{Baibich} and tunnelling 
magnetoresistance~\cite{Julliere,Maekawa,Moodera} in metallic spin valves 
have received much attention~\cite{Wolf}. 
In the latter, on the other hand, the spin-orbit interaction (SOI) plays 
an essential role to connect the charge and spin degrees of freedom. 
However, the role of the SOI in the magnetoresistance has not been considered 
seriously thus far. From this viewpoint, the recently discovered topological 
insulator offers an interesting laboratory to search for the possible 
spintronics functions with the strong SOI.

Recent theoretical and experimental discovery of the two-dimensional 
quantum spin Hall system
\cite{Mele,Bernevig,wu2006,xu2006,Fu,Qi,Bernevig2,Konig},
and its generalization to the topological insulator in three dimensions
\cite{Moore,Fu2,Teo,Hsieh,Qi} have established the new state of matter 
in the time-reversal symmetric systems.
The topological order in the bulk with the gap dictates that
there should be the one-dimensional channels along the edge of 
the two-dimensional sample, or the two-dimensional metal on the surface 
of the three dimensional sample. These edge and surface states are 
protected by the time-reversal symmetry and the topology of the bulk gap, 
and are robust against the disorder scattering and electron-electron interactions.
   
In topological surface state on 3D topological insulator, 
the electrons obey the 2D Dirac equations. 
This corresponds to the infinite mass Rashba model\cite{Rashba}, where only one of the spin-split bands exists.
This has been beautifully demonstrated by the 
spin- and angle-resolved photoemission spectroscopy~\cite{Hasan,Ando}.
Therefore, the next step is to unveil the unique property 
of the surface state of the topological insulators, 
in particular that relevant to magnetism~\cite{Maciejko,Qi2,Liu,Yokoyama,Tanaka}.
One remarkable feature of the Dirac fermions is that the Zeeman 
field acts like vector potential: the Dirac Hamiltonian is transformed 
as ${\bf{k}} \cdot {\bm{\sigma }} \to ({\bf{k}} + {\bf{H}}) \cdot {\bm{\sigma }}$ by the Zeeman field ${\bf{H}}$. Therefore, we can expect anomalous spin related property by the magnetic field in topological insulator. 
This clearly contrasts with the Dirac fermions on graphene since ${\bm{\sigma }}$ is pseudospin there.\cite{Haugen}

\begin{figure}[htb]
\begin{center}
\scalebox{0.8}{
\includegraphics[width=8.5cm,clip]{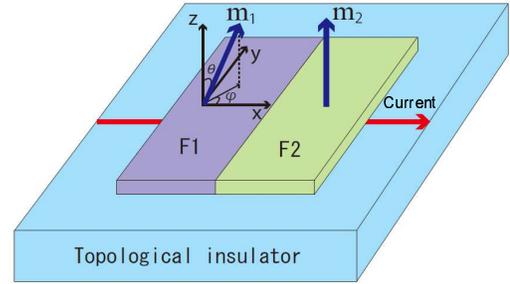}
}
\end{center}
\caption{(Color online) schematics of F1/F2 junction. 
The ferromagnetism is induced in the topological surface state 
due to the proximity effect by the ferromagnetic insulators deposited on the top. The current flows on the surface of the topological insulator. 
}
\label{fig1}
\end{figure}

In this paper, we study charge transport in  2D topological ferromagnet/feromagnet junction. 
The ferromagnet is made of the topological surface with a ferromagnetic insulator on the top. 
We uncover anomalous magnetoresistance in this spin-valve: the conductance strongly depends on  the in-plane rotation with respect to the other magnetization direction. 
Moreover, in sharp contrast to the conventional magnetoresistance effect, the conductance may have its minimum at the parallel configuration,  while it may take a maximum near antiparallel configuration. This is due to the connectivity of the wavefunction across the junction. 


We consider 2D ferromagnet/feromagnet junctions which is abbreviated as F1/F2 in the following. We focus on charge transport at the Fermi level inside the bulk gap of the topological insulator, which is described by the 2D Dirac Hamiltonian
\begin{eqnarray}
H = \left( {\begin{array}{*{20}c}
   {m_z } & {k_x  + m_x  - i(k_y  + m_y )}  \\
   {k_x  + m_x  + i(k_y  + m_y )} & { - m_z }  \\
 \end{array} } \right)
\end{eqnarray}
where $m_x, m_y$ and $m_z$ are exchange field and we set $v_F=\hbar=1$. 
The ferromagnetism is induced due to the proximity 
effect by the ferromagnetic insulators  deposited on the top as shown in Fig. \ref{fig1}. 
The interface is parallel to $y$-axis and located at $x=0$. 
We choose the exchange field in the F1 side as ${\bf{m}}_1$ $=(m_x ,m_y ,m_z ) = 
m_1 (\sin \theta \cos \varphi, \sin \theta \sin \varphi, \cos \theta)$ 
while in the F2 side, we set $m_x=m_y=0$ and $m_z=m_2$. 
In actual experiment, one can use a magnet with very strong easy axis 
anisotropy for F2, and a soft magnet for F1 which can be controlled by 
a weak magnetic field.

We consider the juction between different ferromagnets. 
This type of interface should contain a built-in electric field. Thus, we take into account the potential drop $V$ in F2 which represents the difference of the Fermi energies in the two ferromagnets. Also, due to the mismatch effect, some barrier region may be formed near the interface. We describe this region of the length $L$ by the Dirac fermion with the barrier potentail $U$. 
Note that the potentials $V$ and $U$ may be tunable by gate electrode. 
Then, with the above Hamiltonian, 
wave function in the F1 side is given by 
\begin{widetext}
\begin{equation}
\psi (x \le 0) = \frac{1}{{\sqrt {2E(E - m_z )} }}e^{i k_x x} \left( {\begin{array}{*{20}c}
   {k_x  + m_x  - i(k_y  + m_y )}  \\
   {E - m_z }  \\
\end{array}} \right) + \frac{r}{{\sqrt {2E(E - m_z )} }}e^{ - i k_x x} \left( {\begin{array}{*{20}c}
   { - k_x  - m_x  - i(k_y  + m_y )}  \\
   {E - m_z }  \\
\end{array}} \right)
\end{equation}
\end{widetext}
while the wave function in the barrier region is given by 
\begin{eqnarray}
\psi (0 \le x \le L) = ae^{ik''_x x} \left( {\begin{array}{*{20}c}
   {k''_x  - ik_y }  \\
   {E - U}  \\
\end{array}} \right) + be^{ik''_x x} \left( {\begin{array}{*{20}c}
   {k''_x  - ik_y }  \\
   {E - U}  \\
\end{array}} \right)
\end{eqnarray}
 and that in the F2 side reads 
\begin{eqnarray}
\psi (x \ge L) = \frac{t}{{\sqrt {2E'(E' - m_2)} }}e^{i k'_x x} \left( {\begin{array}{*{20}c}
   {k'_x  - ik_y }  \\
   {E' - m_2}  \\
\end{array}} \right)
\end{eqnarray}
with $E'=E-V$, where $E = \sqrt {m_z ^2  + (k_x  + m_x )^2  + (k_y  + m_y )^2 }  =  - \sqrt {k_x^{\prime \prime 2}  + k_y^2 }  + U =  \pm \sqrt {m_2^2  + k_x^{\prime 2}  + k_y ^2 }  + V$. Here, $\pm$ sign corresponds to the upper and lower bands. Below, "n" and "p" mean that the Fermi level crosses the upper and the lower bands, respectively. Also, $r$ and $t$ are reflection 
and transmission coefficients, respectively. 
It should be noted that the Fermi surface in the F1 is shifted by 
$(-m_x,-m_y)$ from the origin. 
Due to the translational invariance along the $y$-axis, 
the momentum $k_y$ is conserved. Hence, the common 
factor $e^{i k_y y}$ is omitted above.

By matching the wavefunctions at the interface $x=0$ and $L$, we obtain the transmission coefficient $t$. We consider the situation that the barrier region is sufficiently narrow so that we can take the limit of $U \to \infty$ and $L \to 0$ while keeping $Z \equiv UL$ = const.
Here, we omit the expression of $t$ because it is rather complicated but we note that it contains the barrier parameter $Z$ only in the form of $\cos Z$ and $\sin Z$. 
Consequently, the transmission probability and hence the conductance are $\pi$ periodic with respect to $Z$. 
In the presence of $Z$, the spin direction of wavefunction rotates through the barrier region, similar to the spin transister. \cite{Datta}
Thus, with increasing $Z$, the connectivity of the wavefunction changes, which crucially influences the conductance.

We parametrize $k_x  + m_x  = k_F \cos \phi ,k_y  + m_y  = k_F \sin \phi $. 
Then, we have $E = \sqrt {m_z ^2  + (k_x  + m_x )^2  + (k_y  + m_y )^2 }  = \sqrt {m_z ^2  + k_F ^2 }$.

Finally, we obtain the normalized tunneling conductance as 
\begin{equation}
\sigma  = \frac{1}{2}\int_{ - \frac{\pi }{2}}^{\frac{\pi }{2}} 
{d\phi \left| t \right|^2 {\mathop{\rm Re}\nolimits} \left[ {\frac{{k'_x }}{E}} \right]}.
\end{equation}

Now, let us discuss the applicability of our model. 
Typical value of induced exchange field due to the magnetic proximity effect would be 5$\sim$50 meV\cite{Haugen,Chakhalian}, although this depends on the interface property and the material choice of the ferromagnet. On the other hand, $E$ can be tuned by gate electrode or doping below the bulk energy gap ($\sim$ 100 meV). \cite{Zhang}
Due to the presence of the ferromagnet, time reversal symmetry is broken. This would tame the robustness against disoder. However, high quality topological insulator  can be fabricated now, the mean free path of which is sufficiently large,\cite{Hor} and hence localization does not occur in the temperature region of our interest and surface state is stable for exchange field smaller than the bulk energy gap. 
The above approximation of discontinuous change of potential can be justified as follows. The characteristic length of the wavefuction is $\xi  = \hbar v_F/m_z$ while Thomas-Fermi screening length $\lambda$ is given by $1/\lambda  = e^2 N(E)$ in 2D where $N(E)$ is the density of states at the Fermi level.  Then, we have $\xi /\lambda  \sim E/m_z$ using $v_F  \simeq 6 \times 10^5 m/s$ for Bi$_2$Se$_3$ \cite{Zhang}. Thus, we obtain $\xi > \lambda$ for $E>m_z$.

\begin{figure}[htb]
\begin{center}
\scalebox{0.8}{
\includegraphics[width=11.0cm,clip]{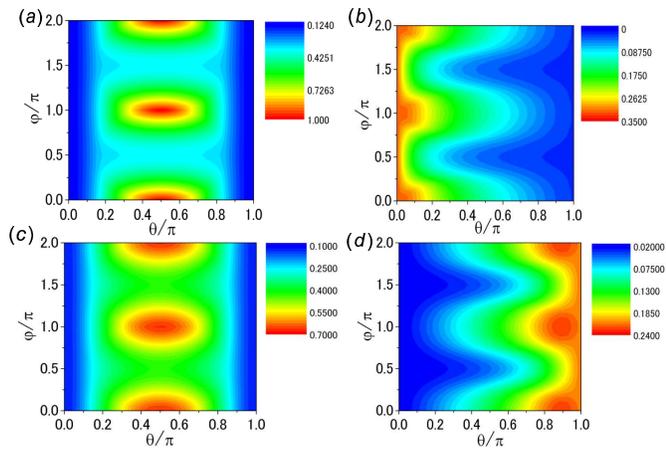}
}
\end{center}
\caption{(Color) tunneling conductance $\sigma$ with $Z=0$ for $m_2=0$ ((a) and (c)), and  $m_2=\sqrt{0.9}E$ ((b) and(d)). n-n junction at $V=0$ in (a) and (b). p-n junction at $V=2E$ in (c) and (d).  }
\label{fig2}
\end{figure}

\begin{figure}[htb]
\begin{center}
\scalebox{0.8}{
\includegraphics[width=11.0cm,clip]{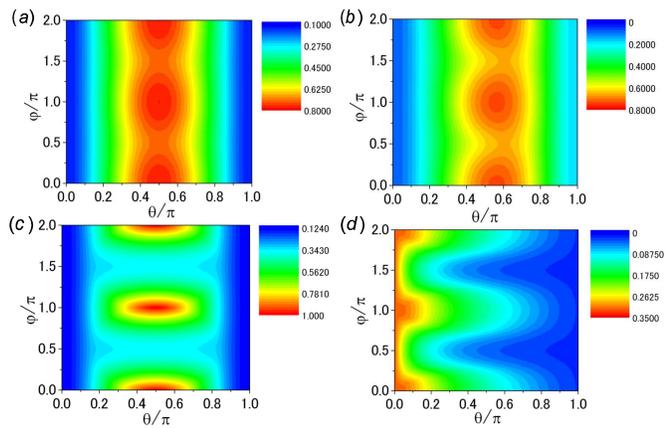}
}
\end{center}
\caption{(Color) similar plots to Fig. \ref{fig2} with $Z=\pi/2$ for $m_2=0$ ((a) and (c)), and  $m_2=\sqrt{0.9}E$ ((b) and(d)). n-n junction at $V=-E$ in (a) and (b). p-n junction at $V=2E$ in (c) and (d). }
\label{fig3}
\end{figure}

In the following, we will show results for $m_1=\sqrt{0.9}E$. 
The tunneling conductance strongly depends on how the wavefunctions connect between both side, which we will explain with Fig. \ref{fig2} for $Z=0$  and Fig. \ref{fig3} for $Z=\pi/2$.
To describe the physics, we first consider the $Z=0$ case. 
 
\begin{figure}[htb]
\begin{center}
\scalebox{0.8}{
\includegraphics[width=7.0cm,clip]{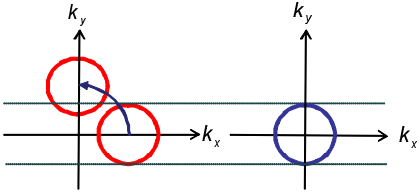}
}
\end{center}
\caption{(Color online)  Positions of Fermi surfaces. On the F1, the Fermi surface moves as illustrated, as ${\bf{m}}_1$ rotates around $z$-axis.}
\label{fig4}
\end{figure}

In Fig. \ref{fig2}, we show the normalized tunneling conductance $\sigma$ in n-n junction for (a) $m_2=0$ and (b) $m_2=\sqrt{0.9}E$. In Fig. \ref{fig2} (a), the F2 is no more ferromagnetic. Nevertheless, the conductance strongly depends on the direction of the magnetization in the F1. 
At $\theta=0$ or $\pi$, the mismatch of the wavefunctions between the two sides and that of the sizes of Fermi surfaces suppresses $\sigma$, because the energy $E$ is near the bottom of the upper band in F1 while there is no gap in F2. 
At $\theta=\pi/2$, on the other hand, the wavefunctions and the sizes of the Fermi surfaces are the same on both sides except the shift of Fermi surface in the momentum space due to the in-plane component of the magnetization as shown in Fig. \ref{fig4}. However, this misfit of the in-plane momentum between the two sides gives rise to a strong dependence of $\sigma$ on the in-plane rotation angle $\varphi$, which is not seen in the conventional magnetoresistance effect.
Since $k_y$ is conserved, 
the positions of the Fermi surfaces strongly influence the charge transport: 
if exchange field points to $x$-axis, there is no evanescent wave. 
On the other hand, when exchange field is applied in $y$-direction, the Fermi 
surface moves to the $k_y$ direction and hence the overlap region 
between $k_y$'s in the F1 and F2 is reduced. Therefore, the number 
of the evanescent modes increases and hence the conductance is strongly suppressed. 
Thus, we can obtain \textit{giant} magnetoresistance in this system. 

In Fig. \ref{fig2} (b), the conductance is large at the parallel configuration ($\theta=0$) while it is small for antiparallel 
configuration ($\theta=\pi$) . 
This $\theta$ dependence, similar to the conventional magnetoresistance effect \cite{Julliere,Maekawa}, can be understood by the overlap intergral of the 
wavefunctions on both sides, as discussed later. 
Note that at the antiparallel configuration,  the domain wall 
structure generates the edge state at the interface. However, 
this edge state is merged with the surface state and do not 
make a significant contribution to the conductance. 

We show tunneling conductance in p-n junction with $V=2 E$ for $m_2=0$ in Fig.\ref{fig2} (c) and  $m_2=\sqrt{0.9}E$ in Fig. \ref{fig2} (d).
In Fig. \ref{fig2} (c), a similar tendency to Figure 
\ref{fig2}(a) is seen. 
In Fig. \ref{fig2} (d), \textit{in stark contrast to the conventional 
magnetoresistance effect, the conductance takes minimum at the parallel 
configuration ($\theta=0$) while it takes maximum  near antiparallel 
configuration ($\theta=\pi$).}

\begin{figure}[htb]
\begin{center}
\scalebox{0.8}{
\includegraphics[width=9.0cm,clip]{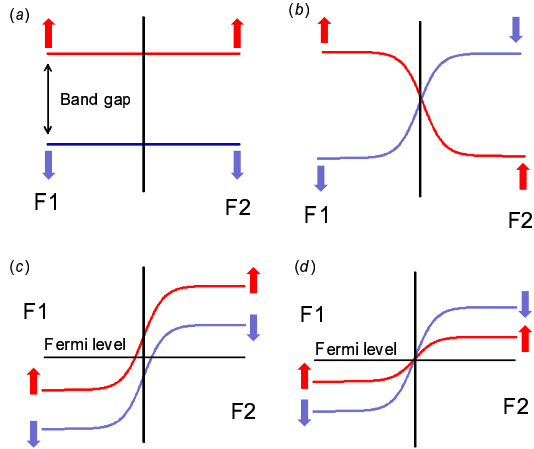}
}
\end{center}
\caption{(Color online) Connectivity of the wavefunction across the n-n junction ((a) and (b)), and p-n junction ((c) and (d)). 
The magnetizations are parallel ($\theta=0$) in (a) and (c), while they are antiparallel ($\theta=\pi$) in (b) and (d). The arrows represent the electron's spin. 
}
\label{fig5}
\end{figure}

To understand these results intuitively, we describe the underlying physics in Fig. \ref{fig5} where the arrows indicate the spin directions in the limiting case of $\left| {m_z } \right| \to \infty $, showing the connection of the wavefunctions on both sides. 
In the n-n junctions,  the tunneling amplitude is determined by the 
overlap of the same eigenfunctions for parallel configuration ($\theta=0$), 
while for antiparallel configuration ($\theta=\pi$) it is given by the overlap 
of the different eigenfunctions, as shown in Figs. \ref{fig5} (a) and (b). 
Thus,  the tunneling amplitude takes its maximum at $\theta=0$, and this 
explains the $\theta$ dependence of the conductance in  Fig. \ref{fig2} (b). 
In a similar way, in p-n junctions, we find that the tunneling amplitude becomes 
larger at  $\theta=\pi$ than  that at $\theta=0$ as shown in Figs. \ref{fig5} (c) and (d).  
This is the origin of the anomalous $\theta$ dependence of the conductance in  Fig. \ref{fig2} (d).


Next, we consider the influence of the barrier potential $Z$ and the potential drop $V$.
Figure \ref{fig3} exhibits tunneling conductance with $Z=\pi/2$ for n-n junction at $V=-E$ ((a) and (b)), and p-n junction at $V=2E$ ((c) and (d)), which should be compared with  Figs. \ref{fig2} (a), (b) and (c), (d), respectively.
In the n-n junction, at $V=-E$ the Fermi surface becomes larger than that at $V=0$. Then, the effect of the shift of the Fermi surface becomes less important. This leads to the weak $\varphi$ dependence as shown in Figs. \ref{fig3} (a) and (b). 
Note that the spin of the eigenstate of Eq.(1) is parallel to $ (k_x  + m_x ,k_y  + m_y ,m_z )^t$.
Therefore, the in-plane component of the spin in the wavefunction in F2 is dominant for $V=-E$. The conductance is largest at $\theta=\pi/2$, when the in-plane spin component in F1 and hence the overlap of the wavefunctions between F1 and F2 are maximum (Figs. \ref{fig3} (a) and (b)).
In Fig. \ref{fig3} (c), since $m_2=0$,  the spin rotation by $Z$ does not make a significant change in the conductance compared to Fig. \ref{fig2} (c).
At $Z=\pi/2$ the spin is half rotated, and therefore the tendency becomes opposite comparing Fig. \ref{fig2} (d) and Fig. \ref{fig3} (d). 


In summary, we studied charge transport in  2D topological ferromagnet/feromagnet junction. 
The ferromagnet is made of the topological surface with a ferromagnetic insulator on the top. 
We found anomalous magnetoresistance in this topological spin-valve.


This work is supported by Grant-in-Aid for Scientific
Research (Grants No. 17071007, 17071005, 19048008 19048015,
and 21244053) from the Ministry of Education, Culture,
Sports, Science and Technology of Japan. T.Y. acknowledges support by JSPS.


\end{document}